\input{aipcheck}

\documentclass[
    ,final            
  ]
  {aipproc}

\layoutstyle{8x11single}

\begin{document}

\title{Time-Reversal Invariance Violation in Heavy and in Few-body Nuclei}


\author{Vladimir Gudkov}{
  address={Department of Physics and Astronomy, University of South Carolina, Columbia, SC, 29208}
}

\author{Young-Ho Song}{
  address={Department of Physics and Astronomy, University of South Carolina, Columbia, SC, 29208}
}

\begin{abstract}
Time reversal invariance violating (TRIV) effects in neutron scattering are very important in a search for new physics, being complementary to neutron and atomic electric dipole moment measurements. In this relation, a sensitivity of  TRIV observables to different models of CP-violation and  their dependencies on nuclear structure, which can lead to new enhancement factors, are discussed.
\end{abstract}

\maketitle



  TRIV in nuclear physics has been a subject of experimental and theoretical investigation  for several decades.
  One important advantage to the search for TRIV in nuclei is the possibility of an enhancement of T-violating observables  by many orders of  magnitude  due to the complex nuclear structure  (see, i.e. paper \cite{Gudkov:1991qg} and references therein). Moreover, the variety of  nuclear systems to measure T-violating parameters provides assurance  that a possible ``accidental'' cancelation of T-violating effects due to unknown structural factors related to the strong interactions in the particular system would be avoided.  Taking into account that different models of the CP-violation may contribute differently to a particular T/CP-observable\footnote{For example, the QCD $\theta$-term can contribute to the neutron EDM but cannot be observed in $K^0$-meson decays. On the other hand, the CP-odd phase of the Cabibbo-Kobayashi-Maskawa matrix was measured in  $K^0$-meson decays, but its contribution to the neutron EDM is extremely small and beyond the reach of the current experimental precision.}, which  may have  unknown theoretical uncertainties, TRIV nuclear effects could be considered  valuable complementary experiments  to  electric dipole moment (EDM) measurements.

One of the  promising approaches to the search of TRIV in nuclear reactions is the measurement of TRIV effects in a transmission of polarized neutrons through a polarized target.
For the observation of TRIV and parity violating (PV) effects,  one can consider  effects related to the $\vec{\sigma}_n\cdot({ \vec{p}}\times{\vec{I}})$ correlation, where  $\vec{\sigma}_n$ is the neutron spin, ${\vec{I}}$ is the target spin,
and $\vec{p}$ is the neutron momentum, which can be observed in the transmission of polarized neutrons through a target with a polarized nuclei.
 This correlation leads to a
difference \cite{Stodolsky:1982tp} between the total neutron cross sections $\Delta\sigma_{\not{T}\not{P}}$  for $\vec{\sigma}_n$
parallel and anti-parallel to ${\vec{p}}\times{\vec{I}}$
and to neutron spin rotation angle \cite{Kabir:1982tp} $\phi_{\not{T}\not{P}}$  around the axis
$\vec{p}\times{\vec{I}}$
\begin{equation}
\label{cc}
\Delta\sigma_{\not{T}\not{P}}=\frac{4\pi}{p}{\rm Im}(f_{+}-f_{-}),\qquad \frac{d\phi_{\not{T}\not{P}}}{dz}=-\frac{2\pi N}{p}{\rm Re}(f_{+}-f_{-}).
\end{equation}
Here, $f_{+,-}$ are the zero-angle scattering amplitudes for neutrons polarized
parallel and anti-parallel to the $\vec{p}\times{\vec{I}}$ axis, respectively;
 $z$ is the target length and $N$ is the number of target nuclei per
unit volume.
The unique feature of these TRIV effects (as well as the similar effects  related to TRIV and parity conserving correlation $\vec{\sigma}_n\cdot({ \vec{p}}\times{\vec{I}})\cdot({ \vec{p}}\cdot{\vec{I}})$) is  the absence of false TRIV effects due to the final state interactions (FSI) (see, for example \cite{Gudkov:1991qg} and references therein), because these effect  are related to   elastic scattering at a zero angle. The general theorem about the absence of FSI for TRIV effects in elastic scattering has been proved first by R. M. Ryndin \cite{Ryndin:fsi} (see, also \cite{Ryndin:1965,Ryndin:1969,Gudkov:1991qg}).  Since this theorem is very important, we give a brief sketch of the proof for the case of the zero angle elastic scattering following \cite{Ryndin:fsi,Gudkov:1991qg}. It is well known that the T-odd angular correlations in scattering and in a particle decay have no relation to TRIV, i.e. they have non-zero values in any process with strong, electromagnetic, and weak interactions. This is because TRI, unlike parity conservation, does not provide a constrain on amplitudes of any process, but rather relates two different processes: for example,  direct and inverse channels of reactions.
However, for the case when the process can be described in the first Born approximation, we can relate T-odd correlations to TRIV interactions. Indeed,  the unitarity condition for the scattering matrix in terms of the reaction matrix $T$, which is proportional to the scattering amplitude, can be written as \cite{Landau:3}
\begin{equation}\label{unit}
T^{\dag}-T=iTT^{\dag}
\end{equation}
The first Born approximation can be used when  the right side of the unitarity equation is much smaller than the left side, and results in hermitian $T$-matrix
\begin{equation}\label{herm}
<i|T|f>=<i|T^*|f>,
\end{equation}
which with TRI condition
\begin{equation}\label{tri}
<f|T|i>=<-i|T|-f>^*
\end{equation}
leads to the constrain on the $T$-matrix as
\begin{equation}\label{todd}
<f|T|i>=<-f|T|-i>^*.
\end{equation}
This condition forbids T-odd angular correlations, as is the case with the P-odd correlations when  parity is conserved. (Here the minus signs in matrix elements mean the opposite signs for particle spins and momenta in the corresponding states.) For the case of the zero angle elastic scattering, the initial and final states  coincide ($i=f$), and when combined with TRI condition (\ref{tri}), result in  Eq.(\ref{todd}) without the violation of unitarity (\ref{herm}). Therefore, in this case, FSI cannot mimic  T-odd correlations, which  originated from TRIV interactions. Then, an observation of a non-zero value of  TRIV effects in neutron transmission directly indicates  TRIV, exactly like in the case of neutron EDM \cite{Landau:1957}.

Moreover,  these TRIV effects  are  enhanced \cite{Bunakov:1982is} by a factor of about $10^6$ in neutron induced  nuclear reactions (the similar  enhancement was observed for PV effects related to  $(\vec{\sigma}_n\cdot{ \vec{p}})$ correlation in neutron transmission through nuclear targets). Since TRIV and PV effects have  similar enhancement factors,  it is convenient to consider \cite{Gudkov:1990tb,Gudkov:1991qg} the ratio $\lambda$ of TRIV to PV effect at the same nuclei and at the same neutron energy as the measure of TRIV effect, because for this ratio, most nuclear structure effects  cancel each other out.  As a result, one can  estimate $\lambda \sim g_{T}/g_{P}$, where $g_{T}$ and $g_{P}$ are TRIV and PV nucleon nucleon coupling constants.  Theoretical predictions for $\lambda$ are varying from $10^{-2}$ to $10^{-10}$ for different models of the CP violation (see, for example, \cite{Herczeg:1987gp,Gudkov:1990tb,Gudkov:1991tp,Gudkov:1992yc,Gudkov:1995tp} and references therein). Therefore,  one can estimate a range of  possible values of the TRIV observable and relate a particular mechanism of the CP-violation to their values. These estimates show that these effects could be measured at the new  spallation neutron facilities, such as the SNS at the Oak Ridge National Laboratory or the J-SNS at J-PARC in Japan.
However,  existing estimates of CP-violating effects for nuclear reactions have at most  order of  magnitude of accuracy.
In this relation, it is desirable to compare
the calculation of TRIV effects for complex nuclei to
the  calculations of the same effects for the simplest few-body systems and to the calculations of EDMs.

Using the results of the recent calculations of PV and TRIV effects in neutron deuteron scattering \cite{Song:2010sz,Song:2011jh,Song:2011sw}, one can calculate the parameter $\lambda$ for this  reaction and  compare it to the case of the complex nuclei. Let us consider the ratio of the TRIV difference of total cross sections in  Eq.(\ref{cc})  given in \cite{Song:2011jh}
\begin{eqnarray}
\label{eq:PTP}
P_{\not{T}\not{P}}=\frac{\Delta\sigma_{PT}}{2\sigma_{tot}}&=&
\frac{(-0.185 \mbox{ b})}{2\sigma_{tot}}
[\bar{g}_\pi^{(0)}+0.26 \bar{g}_\pi^{(1)}
-0.0012 \bar{g}_\eta^{(0)}+0.0034 \bar{g}_\eta^{(1)} \\ \nonumber
&-&0.0071 \bar{g}_\rho^{(0)}+0.0035 \bar{g}_\rho^{(1)}
+0.0019 \bar{g}_\omega^{(0)}-0.00063 \bar{g}_\omega^{(1)}]
\end{eqnarray}
to the corresponding PV difference \cite{Song:2010sz}
\begin{eqnarray}
  \label{eq:PP}
P_{\not{P}}=\frac{\Delta\sigma_P}{2\sigma_{tot}}&=&
  \frac{(0.395 \mbox{ b})}{2\sigma_{tot}}\left[
  h_\pi^1+h_\rho^0(0.021)+h_\rho^1(0.0027)
         +h_\omega^0(0.022)+h_\omega^1(-0.043)
         +h_\rho^{'1}(-0.012)
  \right].
\end{eqnarray}
Here, we use one meson exchange model, known as the DDH model for PV nucleon interactions, to calculate both effects; in the above expressions, $\bar{g}$ and $h$ are meson- nucleon TRIV and PV coupling constants, correspondingly (see for details \cite{Song:2010sz,Song:2011jh}). From these expressions, one can see that  contributions from the pion exchange are dominant for both TRIV and PV parameters. Then, taking into account only the dominant pion meson contributions, one can estimate $\lambda$ as
\begin{equation}
\label{lambda}
\lambda=\frac{\Delta\sigma_{\not{T}\not{P}}}{\Delta\sigma_{\not{P}}} \simeq (-0.47)\left(\frac{\bar{g}^{(0)}_\pi}{h_\pi^1}
                   +(0.26)\frac{\bar{g}^{(1)}_\pi}{h_\pi^1}\right),
\end{equation}
which is in a good agreement with the estimate for the complex nuclei  \cite{Gudkov:1990tb}.

Also, we can relate the obtained parameter $\lambda$ to the existing experimental constrains obtained from EDM measurements, even though the relationships are model dependent.  For example, the CP-odd coupling constant $\bar{g}^{(0)}_\pi$ could be related to the value of the neutron EDM $d_n$ generated via a $\pi$-loop in the chiral limit \cite{Pospelov:2005pr}.  Then, using the experimental limit \cite{Baker:2006ts} on  $d_n$, one can estimate $\bar{g}^{(0)}_\pi $ as less than $2.5 \times 10^{-10}$. The constant $\bar{g}^{(1)}_\pi$ can be bounded using the constraint \cite{Romalis:2000mg} on the $^{199}Hg$ atomic EDM as $\bar{g}^{(1)}_\pi<0.5\times 10^{-11}$ \cite{Dmitriev:2003hs}.

The comparison of the $\lambda$ parameter  with the constrains on the coupling constants from the EDM experiments gives us the opportunity to estimate the ``discovery potential'' for neutron scattering experiments as a possible factor for improving the current limits of the EDM experiments. Then, taking the DDH ``best value'' of $h_\pi^1\sim 4.6\cdot 10^{-7}$,  nuclear enhancement factors, and assuming that the parameter $\lambda$ could be measured with an accuracy of $10^{-5}$ on the complex nuclei, one can see from Eq.(\ref{lambda}) that  the existing limits on the TRIV coupling constants could be improved by  two orders of magnitude. It should be noted that to obtain Eq.(\ref{lambda}), the assumption was made that the $\pi$-meson exchange contribution is dominant for PV effects.  However, there is an indication \cite{Bowman} that the PV coupling constant  $h_\pi^1$ is much smaller than the ``best value'' of the DDH.  Should it be confirmed by the $\overrightarrow{n}+p \rightarrow d+\gamma$ experiment, the  estimate for the sensitivity of $\lambda$ to the TRIV coupling constant may be increased up to two orders of magnitude, as can be seen from Eqs.(\ref{eq:PTP}-\ref{lambda}). This  might increase the relative values of TRIV effects by two orders of magnitude, and as a consequence, the discovery potential of the TRIV experiments could be about $10^4$.

To conclude, the TRIV effects in neutron transmission through a nuclei target are very unique TRIV observables being free from FSI, and are of the same quality as the EDM experiments. These TRIV effects are enhanced by about $10^6$ due to the nuclear enhancement factor. In addition to this enhancement, they might be  structurally enhanced by about $10^2$ if PV $\pi$-nucleon coupling constant is less than  the ``best value'' DDH estimate. Therefore, these types of experiments have a discovery potential of about $10^2 - 10^4$ for the improvement of the current limits on the TRIV interaction obtained from the EDM experiments.
\begin{theacknowledgments}
This work was supported by the DOE grants no. DE-FG02-09ER41621.
\end{theacknowledgments}
\bibliographystyle{aipproc}  
\bibliography{TViolation}

\end{document}